\documentclass[journal]{IEEEtran}
\ifCLASSINFOpdf
 \usepackage[pdftex]{graphicx}

\fi
\usepackage{cite}
\usepackage{graphics}
\usepackage{graphicx}
\usepackage{color}
\usepackage{multirow}  
\usepackage{amsmath}
\usepackage{multicol}
\usepackage{algorithmic}
\usepackage{array}
\usepackage{stfloats}

\usepackage{amsfonts,amssymb}

\par
\begin{document}

%
\title{ Hyperspectral image reconstruction for spectral camera based on ghost imaging via sparsity constraints using V-DUnet}
%
%
%

        

\author{Ziyan~Chen,
        Zhentao~Liu,
        Chenyu~Hu, 
         Heng~Wu,   
        Jianrong~Wu,
         Jinda~Lin, 
          Zhishen~Tong,
           Hong~Yu,    
        and Shensheng~Han
\thanks{ \emph{(Corresponding
author: Zhentao Liu.)}}      
        
\thanks{Ziyan Chen, Zhentao Liu , Jianrong Wu , Jinda Lin and  Zhishen Tong are with the Key Laboratory for Quantum Optics of CAS, Shanghai Institute of Optics and Fine Mechanics, 
Chinese Academy of Sciences, Shanghai, 201800, China  and  with Center of Materials Science and Optoelectronics Engineering, University of Chinese Academy of Sciences, Beijing, 100049, China.
}      
\thanks{Chenyu Hu is with Hangzhou Institute for Advanced Study, University of Chinese Academy of Sciences, Hangzhou, 310024, China.}   
\thanks{Heng Wu is with Guangdong Provincial Key Laboratory of Cyber-Physical System, School of Automation, Guangdong University of Technology, Guangzhou, 510006, China.}
\thanks{ Shensheng Han and Hong Yu are with the Key Laboratory for Quantum Optics of CAS, Shanghai Institute of Optics and Fine Mechanics, 
Chinese Academy of Sciences, Shanghai, 201800, China  and  with Center of Materials Science and Optoelectronics Engineering, University of Chinese Academy of Sciences, Beijing, 100049, China  
and with  Hangzhou Institute for Advanced Study, University of Chinese Academy of Sciences, Hangzhou, 310024, China .}}

\maketitle

\begin{abstract}

Spectral camera based on ghost imaging via sparsity constraints (GISC spectral camera) obtains three-dimensional (3D) hyperspectral information with two-dimensional (2D) compressive measurements in a single shot, which has attracted much attention in recent years. However, its imaging quality and real-time performance of reconstruction still need to be further improved. Recently, deep learning has shown great potential in improving the reconstruction quality and reconstruction speed for computational imaging. When applying deep learning into GISC spectral camera, there are several challenges need to be solved: 1) how to deal with the large amount of 3D hyperspectral data, 2) how to reduce the influence caused by the uncertainty of the random reference measurements, 3) how to improve the reconstructed image quality as far as possible. In this paper, we present an end-to-end V-DUnet for the reconstruction of 3D hyperspectral data in GISC spectral camera. To reduce the influence caused by the uncertainty of the measurement matrix and enhance the reconstructed image quality, both differential ghost imaging results and the detected measurements are sent into the network’s inputs. Compared with compressive sensing algorithm, such as PICHCS and TwIST, it not only significantly improves the imaging quality with high noise immunity, but also speeds up the reconstruction time by more than two orders of magnitude. 

\end{abstract}


%
\IEEEpeerreviewmaketitle

\section{Introduction} 
\IEEEPARstart{G}{HOST} imaging (GI) obtains the image information through intensity correlation of optical fields between the object path and the reference path \cite{Strekalov1995, Cheng2004, Gatti2004, kolobov2007quantum, Shih2008, Shapiro2012}. It can restore the high-dimensional information from the low-dimensional detecting measurements by encoding the image information into the intensity fluctuations of light fields, thus providing a new solution for high dimensional image sensing \cite{Duarte2008, Zhao2012, Hardy2013, Sun2013}. As a typical case, spectral camera based on ghost imaging via sparsity constraints (GISC spectral camera) modulates the 3D hyperspectral information into a 2D spatial intensity fluctuations of light fields, which enables capturing the 3D hyperspectral image information in a single shot \cite{Han2018, Liu2016}. Combined with compressive sensing \cite{Donoho2006, Candes2006, Eldar2009}, it can realize compressive sensing of the information during the acquisition process with improved efficiency. However, the image reconstruction process is full of challenges. Conventional GI reconstruction algorithms, such as differential GI (DGI) \cite{Ferri2010}, suffer from the low reconstruction quality in the case of low sampling rate and low signal to noise ratio. Though Compressive sensing algorithms can contribute to obtain higher reconstruction quality by utilizing prior information of the object, the time-consuming interactive process makes it difficult to reconstruct the image in real time. With recent explosive growth of artificial intelligence, deep learning (DL) has provided new opportunities and tools for computational imaging \cite{Barbastathis2020, Wang2020, Yuan2021, Miao2019, Wang2019, Zhang2019, Wang2019a, Zhang2021, Zhang2020, Sharma2020}. In recent years, DL has also been applied in ghost imaging and has achieved good performance \cite{Wu2020, Zhu2020, Hu2020, Wang2019b, Lyu2017, He2018, Li2020}. Many excellent works set the detected measurements as the net input \cite{Wu2020, Wang2019b, Li2020}, and the sufficient sampling rate for high quality image goes down to a cheerful level. However, these works require that the measurement matrix must be the same during the training and imaging process. Zhu \cite{Zhu2020}  proposes a novel dynamic decoding deep learning framework called Y-net, which introduces the statistical characteristics of the random reference measurements into the net and works well under both fixed and unfixed measurement matrix. Hu \cite{Hu2020} and Lyu \cite{Lyu2017} have also reduced the sensibility of the measurement matrix by setting the conventional ghost imaging results as the network’s input. 

\indent Compared to the 2D reconstruction in GI, introducing deep learning into the reconstruction of 3D hyperspectral information in GISC spectral camera faces the following challenges. Firstly, large-size data need to be processed due to its high dimensional property. Secondly, how to reduce the sensibility of the random reference measurements also plays an important role in the generalization ability of the network. What’s more, the reconstruction quality of 3D hyperspectral information has also to be ensured. In this paper, we propose an end-to-end V-DUnet to reconstruct 3D hyperspectral images of GISC spectral camera. Owing to the encoder and decoder architecture of the Unet \cite{Ronneberger2015}, it can effectively deal with large-size data. And by setting both differential ghost imaging results and the detected measurements as network’s input, V-DUnet has not only successfully reduced the influence caused by the uncertainty of the random reference measurements, but also improved the reconstruction quality of 3D hyperspectral images in GISC spectral camera.

\begin{figure*}[t!]
\centering
\includegraphics[width=14cm,angle=0]{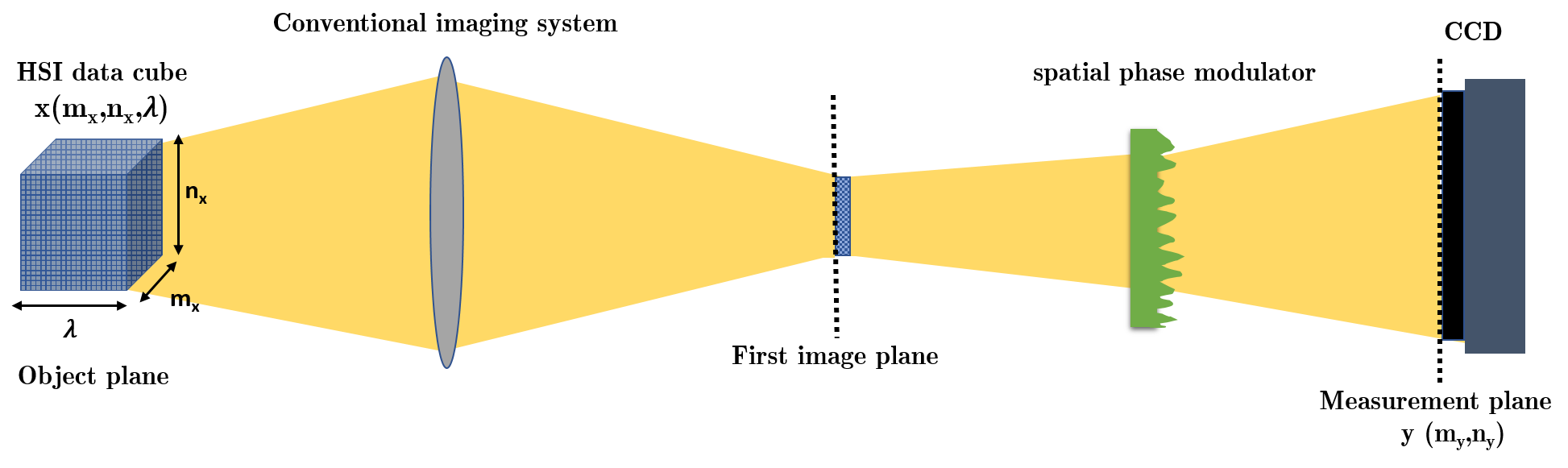}

\caption{The schematic of GISC spectral camera. The system is composed of three modules: (1) A front imaging module (a conventional imaging system), which projects the 3D hyperspectral data cube {\bf x}$(m_x, n_x, \lambda)$ onto the first imaging plane, (2) Modulation module (a spatial random phase modulator), which modulates the light fields in the first imaging plane, (3) Detection module (CCD), which records the speckle patterns in the measurement plane {\bf y}$ (m_y,n_y)$.}
\label{f1}
\end{figure*}

\begin{figure*}[htpb]
\centering
\includegraphics[width=16cm,height=8cm]{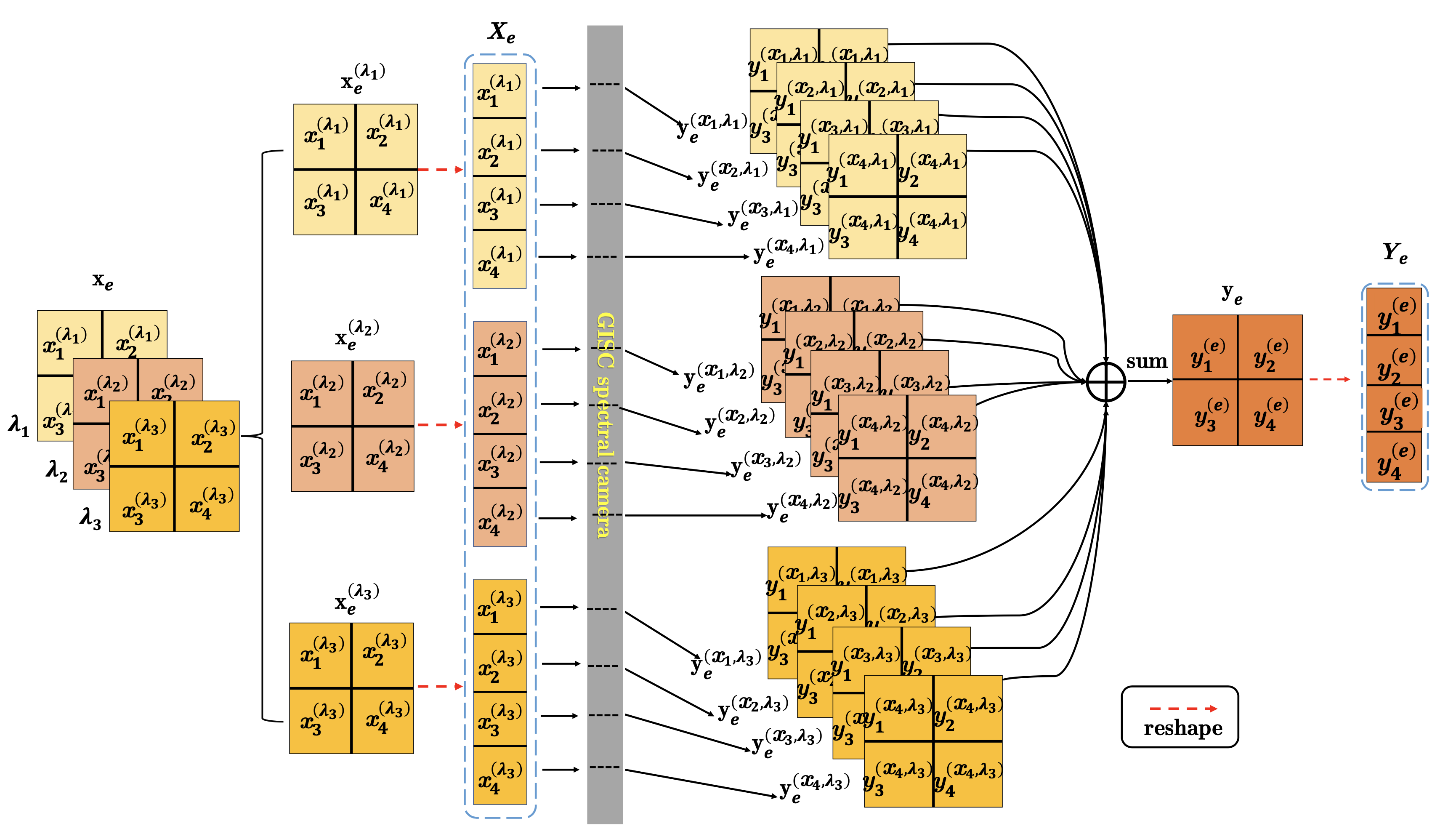}
\caption{An illustration of a tiny HSI data's flow in GISC spectral camera. Each pixel in data cube $\bf x_e$ ($\bf x_e$ have total 12 pixels, they are $x_1^{(\lambda_1)}$, $x_2^{(\lambda_1)}$, $x_3^{(\lambda_1)}$, $x_4^{(\lambda_1)}$, $x_1^{(\lambda_2)}$, $x_2^{(\lambda_2)}$, $x_3^{(\lambda_2)}$, $x_4^{(\lambda_2)}$, and $x_1^{(\lambda_3)}$, $x_2^{(\lambda_3)}$, $x_3^{(\lambda_3)}$, $x_4^{(\lambda_3)}$ respectively) contribute to a corresponding random speckle pattern {( $ {\bf{y_e}}^{(x_1,\lambda_1)}$, $ {\bf{y_e}}^{(x_2,\lambda_1)}$, $ {\bf{y_e}}^{(x_3,\lambda_1)}$, $ {\bf{y_e}}^{(x_4,\lambda_1)}$, $ {\bf{y_e}}^{(x_1,\lambda_2)}$, $ {\bf{y_e}}^{(x_2,\lambda_2)}$, $ {\bf{y_e}}^{(x_3,\lambda_2)}$, $ {\bf{y_e}}^{(x_4,\lambda_2)}$, and $ {\bf{y_e}}^{(x_1,\lambda_3)}$, $ {\bf{y_e}}^{(x_2,\lambda_3)}$, $ {\bf{y_e}}^{(x_3,\lambda_3)}$, $ {\bf{y_e}}^{(x_4,\lambda_3)}$}, respectively) on the CCD detector plane. The detector captures the intensity $ {\bf{y_e}}$ by integrating the total $12$ random speckle patterns.}
 \label{f11}
\end{figure*}

\section{System of GISC spectral camera}

Fig.\ref{f1} shows the schematic of GISC spectral camera. Lights from the 3D hyperspectral image (HSI) ${\bf x} (m_x,n_x,\lambda$) are collected by a conventional imaging system in the first imaging plane and then is modulated by a spatial random phase modulator, finally, the modulated imaging speckle patterns ${\bf y}(m_y,n_y)$ are recorded by an CCD detector (each pixel in the CCD collects the intensity signal from the whole 3D hyperspectral imaging). 
In addition, before the imaging process, the calibrated speckle patterns are pre-determined by scanning long the spatial and spectral dimensions with a monochromatic point source on the object plane. Thus, 3D hyperspectral images can be obtained by calculating the intensity correlation between the calibrated speckle patterns and imaging speckle patterns \cite{Liu2016}. Meanwhile, the imaging process can be written into a matrix form as \cite{Han2018}
\begin{eqnarray}
{Y} =\Phi {X} + \epsilon,
 \label{eq1}
\end{eqnarray}

in which ${X}  \in \mathbb{R} ^{M_xN_xL}$ is reshaped from the HSI data cube ${\bf x}(m_x, n_x, \lambda) \in \mathbb{R} ^{M_x \times N_x \times L}$ where $1 \leqslant m_x \leqslant M_x$, $1 \leqslant n_x \leqslant N_x$ and $1 \leqslant \lambda \leqslant L$, ${Y}  \in \mathbb{R} ^{M_yN_y} $ is reshaped from the measurement image ${\bf y}(m_y, n_y)\in \mathbb{R} ^{M_y \times N_y}$  where $1 \leqslant m_y \leqslant M_y$ and $1 \leqslant n_y \leqslant N_y$ in the CCD detector. $\epsilon$ represents the noise of the system. The pre-determined random measurement matrix $\Phi \in \mathbb{R} ^{M_yN_y \times M_xN_xL  }$ is obtained after $M_x N_x L$ calibration measurements, each
column vector  in $\Phi $ presents a calibrated speckle intensity pattern corresponding to one pixel in HSI. 
\begin{figure*}[t!]
\centering
\includegraphics[width=16cm,height=9cm]{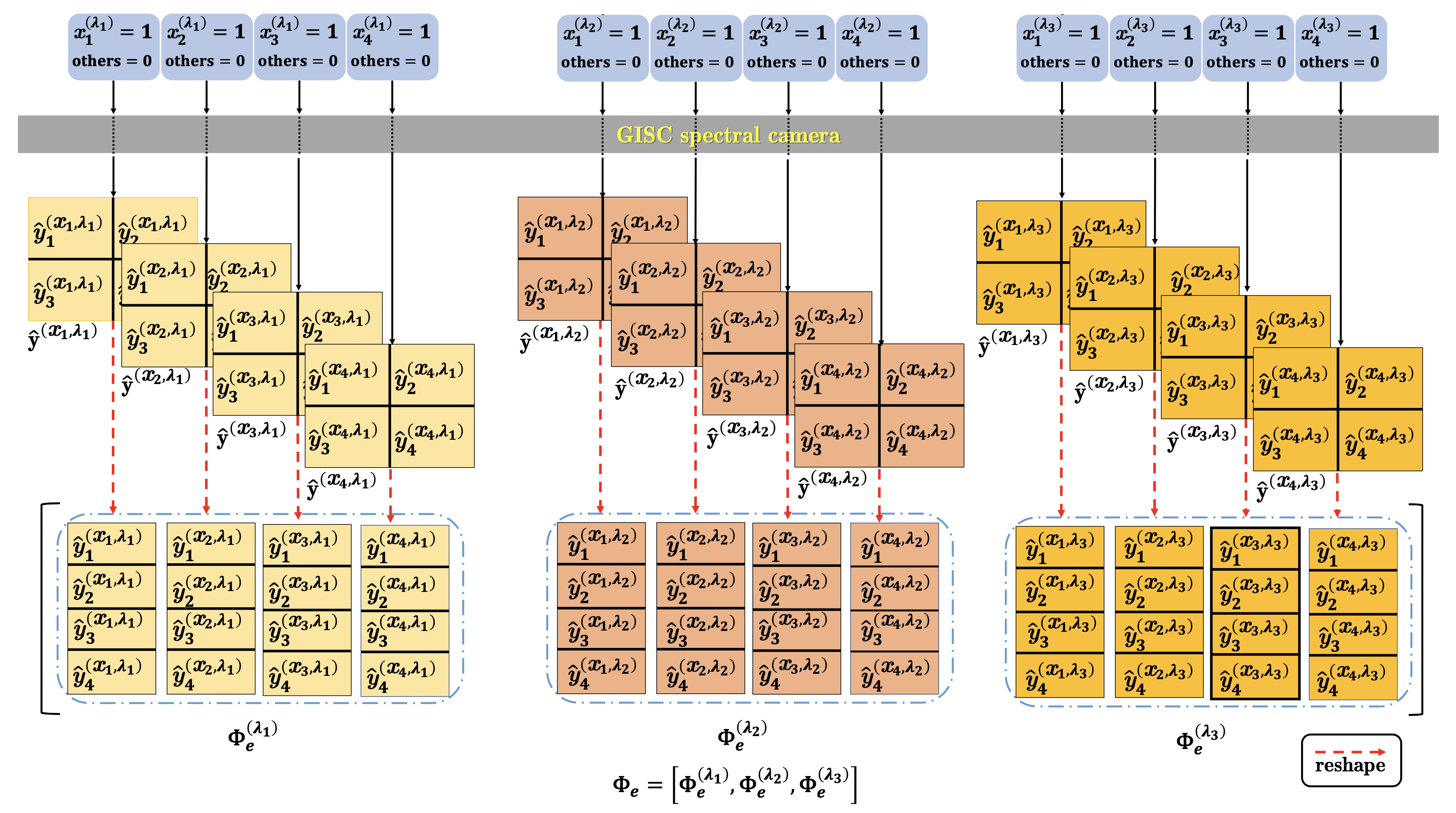}
\caption{Structure of the matrix ${\Phi_e}$ for $M_x=2, N_x=2$, $L=3$ and $M_y=2, N_y=2$. }
\label{f111}
\end{figure*}

In order to have an intuitive view of our GISC spectral camera sensing matrix $\Phi$,  we choose a tiny HSI data cube $\bf {x_e} \in \mathbb{R} ^{2 \times 2 \times 3}$ as an example and set the $\bf {y_e} \in \mathbb{R} ^{2 \times 2}$ to give an illustration. What's more, we suppose the system is noise-clean for simplicity. First, the tiny HSI data's flow in GISC spectral camera is particularly illustrated in Fig.\ref{f11}, each pixel in HSI data cube $\bf x_e$ produces a random speckle pattern on the CCD plane after the interaction of the conventional imaging system and the spatial random phase modulator. In our selected tiny HSI data cube $\bf x_e$, it has total $12$ ($M_x=2$,$N_x=2$ and $L=3$, $2 \times 2 \times 3=12$) pixels $x_1^{(\lambda_1)}$, $x_2^{(\lambda_1)}$, $x_3^{(\lambda_1)}$, $x_4^{(\lambda_1)}$, $x_1^{(\lambda_2)}$, $x_2^{(\lambda_2)}$, $x_3^{(\lambda_2)}$, $x_4^{(\lambda_2)}$, and $x_1^{(\lambda_3)}$, $x_2^{(\lambda_3)}$, $x_3^{(\lambda_3)}$, $x_4^{(\lambda_3)}$, thus the corresponding 12 random speckle patterns are  {$ {\bf{y_e}}^{(x_1,\lambda_1)}$, $ {\bf{y_e}}^{(x_2,\lambda_1)}$, $ {\bf{y_e}}^{(x_3,\lambda_1)}$, $ {\bf{y_e}}^{(x_4,\lambda_1)}$, $ {\bf{y_e}}^{(x_1,\lambda_2)}$, $ {\bf{y_e}}^{(x_2,\lambda_2)}$, $ {\bf{y_e}}^{(x_3,\lambda_2)}$, $ {\bf{y_e}}^{(x_4,\lambda_2)}$, and $ {\bf{y_e}}^{(x_1,\lambda_3)}$, $ {\bf{y_e}}^{(x_2,\lambda_3)}$, $ {\bf{y_e}}^{(x_3,\lambda_3)}$, $ {\bf{y_e}}^{(x_4,\lambda_3)}$}, respectively. $ {\bf{y_e}}$ is the superposition of those total 12 random speckle patterns, namely
{\begin{eqnarray}
  & \bf{y_e}=& {\bf{y_e}}^{(x_1,\lambda_1)}+  {\bf{y_e}}^{(x_2,\lambda_1)} + {\bf{y_e}}^{(x_3,\lambda_1)} + {\bf{y_e}}^{(x_4,\lambda_1)}                      \nonumber\\
             &&  +  {\bf{y_e}}^{(x_1,\lambda_2)}+{\bf{y_e}}^{(x_2,\lambda_2)} + {\bf{y_e}}^{(x_3,\lambda_2)} +{\bf{y_e}}^{(x_4,\lambda_2)}   \nonumber\\
             &&  + {\bf{y_e}}^{(x_1,\lambda_3)}+  {\bf{y_e}}^{(x_2,\lambda_3)} +  {\bf{y_e}}^{(x_3,\lambda_3)} + {\bf{y_e}}^{(x_4,\lambda_3)} 
\label{eq11}
\end{eqnarray} 
}

 \begin{figure*}[htpb]
\normalsize
 \begin{eqnarray}
{\Phi_e} &=& \left[  \Phi_e^{(\lambda_1)}  \ \  \Phi_e^{(\lambda_2)}  \ \  \Phi_e^{(\lambda_3)}   \right]  \nonumber \\
          &=&\left[ \begin{array} {cccccccccccc}
                   {\hat{y}_1^{(x_1,\lambda_1)}} \ {\hat{y}_1^{(x_2,\lambda_1)}} \ {\hat{y}_1^{(x_3,\lambda_1)}} \ {\hat{y}_1^{(x_4,\lambda_1)}} \ 
                   {\hat{y}_1^{(x_1,\lambda_2)}} \ {\hat{y}_1^{(x_2,\lambda_2)}} \ {\hat{y}_1^{(x_3,\lambda_2)}} \ {\hat{y}_1^{(x_4,\lambda_2)}} \
                    {\hat{y}_1^{(x_1,\lambda_3)}} \ {\hat{y}_1^{(x_2,\lambda_3)}} \ {\hat{y}_1^{(x_3,\lambda_3)}}   \ {\hat{y}_1^{(x_4,\lambda_3)}} \\                  
                                     {\hat{y}_2^{(x_1,\lambda_1)}} \ {\hat{y}_2^{(x_2,\lambda_1)}} \ {\hat{y}_2^{(x_3,\lambda_1)}}  \ {\hat{y}_2^{(x_4,\lambda_1)}}  \ {\hat{y}_2^{(x_1,\lambda_2)}} \ {\hat{y}_2^{(x_2,\lambda_2)}} \ {\hat{y}_2^{(x_3,\lambda_2)}} \ {\hat{y}_2^{(x_4,\lambda_2)}}\
                    {\hat{y}_2^{(x_1,\lambda_3)}} \ {\hat{y}_2^{(x_2,\lambda_3)}} \ {\hat{y}_2^{(x_3,\lambda_3)}} \ {\hat{y}_2^{(x_4,\lambda_3)}}      \\
                   
                                       {\hat{y}_3^{(x_1,\lambda_1)}} \ {\hat{y}_3^{(x_2,\lambda_1)}} \ {\hat{y}_3^{(x_3,\lambda_1)}}                                 \ {\hat{y}_3^{(x_4,\lambda_1)}}  \ 
                   {\hat{y}_3^{(x_1,\lambda_2)}} \ {\hat{y}_3^{(x_2,\lambda_2)}} \ {\hat{y}_3^{(x_3,\lambda_2)}}   \ {\hat{y}_3^{(x_4,\lambda_2)}} \
                    {\hat{y}_3^{(x_1,\lambda_3)}} \ {\hat{y}_3^{(x_2,\lambda_3)}} \ {\hat{y}_3^{(x_3,\lambda_3)}}   \ {\hat{y}_3^{(x_4,\lambda_3)}}      \\   
                   
                                                        {\hat{y}_4^{(x_1,\lambda_1)}} \ {\hat{y}_4^{(x_2,\lambda_1)}} \ {\hat{y}_4^{(x_3,\lambda_1)}}                                 \ {\hat{y}_4^{(x_4,\lambda_1)}}  \ 
                   {\hat{y}_4^{(x_1,\lambda_2)}} \ {\hat{y}_4^{(x_2,\lambda_2)}} \ {\hat{y}_4^{(x_3,\lambda_2)}}   \ {\hat{y}_4^{(x_4,\lambda_2)}} \
                    {\hat{y}_4^{(x_1,\lambda_3)}} \ {\hat{y}_4^{(x_2,\lambda_3)}} \ {\hat{y}_4^{(x_3,\lambda_3)}}   \ {\hat{y}_4^{(x_4,\lambda_3)}}      \\    
                                  \end{array} \right]     
\label{eq111}
\end{eqnarray}  
\hrulefill
\end{figure*}

    Second, the calibration measurement process of the sensing matrix $\Phi_e \in \mathbb{R} ^{4 \times 12}$ is displayed in Fig.\ref{f111}. To obtain the sensing matrix $\Phi_e$, one just needs to set the values of each pixel in HSI data cube $\bf x_e$ to $1$ in sequence. As the same data flow process illustrated in Fig.\ref{f11}, $12$ corresponding random speckle patterns $\hat{\bf y}^{(x_1,\lambda_1)}$, $\hat{\bf y}^{(x_2,\lambda_1)}$, $\hat{\bf y}^{(x_3,\lambda_1)}$, $\hat{\bf y}^{(x_4,\lambda_1)}$, $\hat{\bf y}^{(x_1,\lambda_2)}$, $\hat{\bf y}^{(x_2,\lambda_2)}$, $\hat{\bf y}^{(x_3,\lambda_2)}$, $\hat{\bf y}^{(x_4,\lambda_2)}$, and $\hat{\bf y}^{(x_1,\lambda_3)}$, $\hat{\bf y}^{(x_2,\lambda_3)}$, $\hat{\bf y}^{(x_3,\lambda_3)}$, $\hat{\bf y}^{(x_4,\lambda_3)}$ are generated, respectively. And the sensing matrix $\Phi_e$ is finally obtained by reshaping all those patterns to column vectors and placing them in order, as is shown in Fig.\ref{f111} and Eq.\ref{eq111}. Finally, we let $X_e \in \mathbb{R} ^{12}$ represent the column vector reshaped from $\bf x_e$, $Y_e \in \mathbb{R} ^{4}$ represent the column vector reshaped from $\bf y_e$,  thus the formula between $X_e$ and $Y_e$ can be written as    
\begin{eqnarray}
{Y_e} =\Phi_e {X_e},
 \label{eq1111}
\end{eqnarray}    
in which $Y_e=[y_1^{(e)}\ y_2^{(e)}\ y_3^{(e)}\ y_4^{(e)}]^T$, $X_e$=$ [ x_1^{(\lambda_1)}$ \ $x_2^{(\lambda_1)}$ \ $x_3^{(\lambda_1)}$ \ $x_4^{(\lambda_1)}$ \ $x_1^{(\lambda_2)}$ \  $x_2^{(\lambda_2)}$ \ $x_3^{(\lambda_2)}$ \ $x_4^{(\lambda_2)}$ \ $x_1^{(\lambda_3)}$ \ $x_2^{(\lambda_3)}$ \ $x_3^{(\lambda_3)}$ \ $x_4^{(\lambda_3)} ]^T$.

\section{The proposed framework}
\begin{figure*}[t!]
\centering
\includegraphics[width=18cm,angle=0]{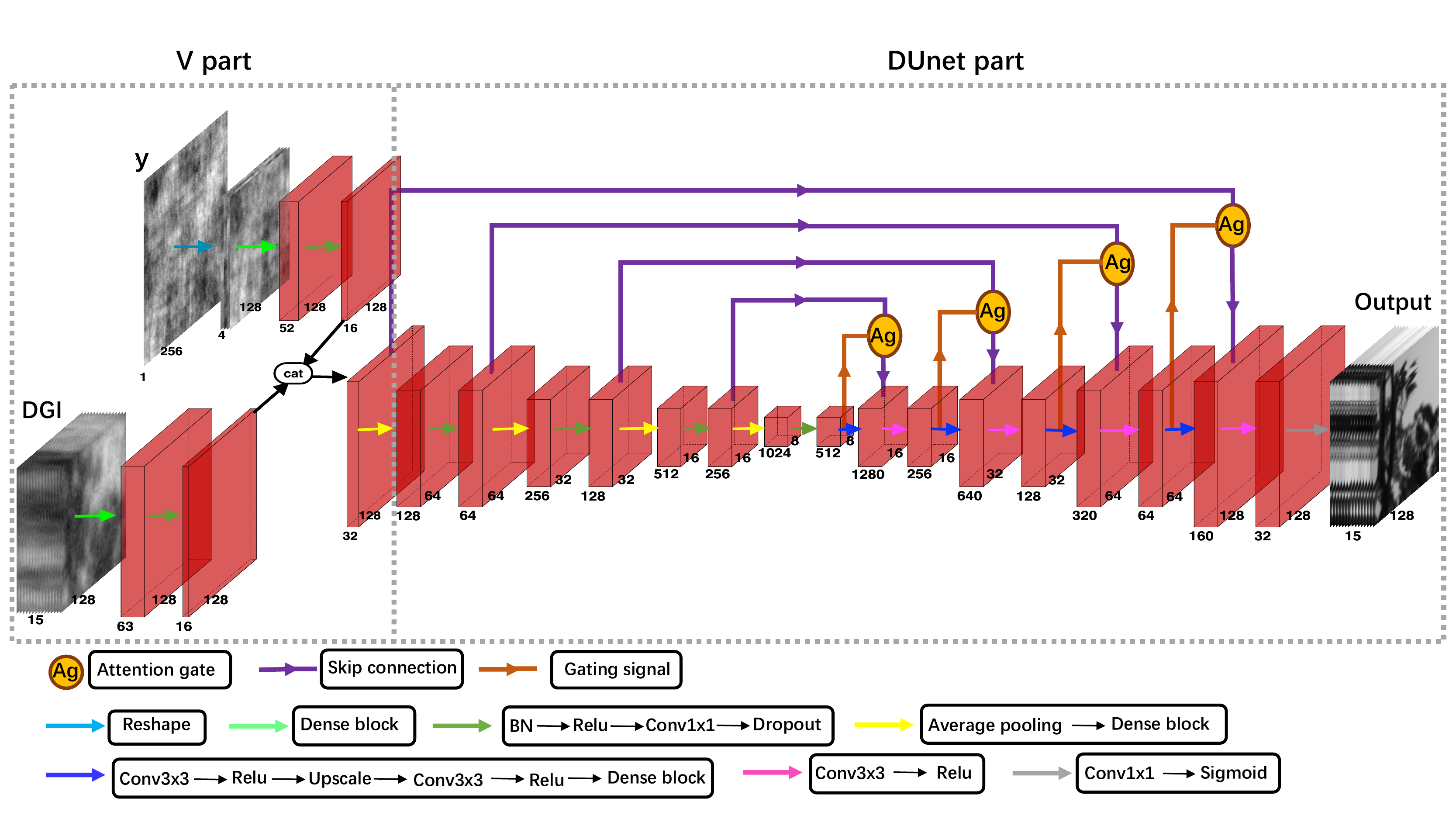}
\caption{The architecture of the proposed V-DUnet. BN, batch normalization; Conv $3 \times 3$, convolution with filter size $3 \times 3$; Conv $1 \times 1$, convolution with filter size $1 \times1$; Dropout, dropout rate is $0.05$; Relu, rectified linear unit;  Average pooling , stride ($2, 2$); Upscale, factor $2$.}
\label{f2}
\end{figure*}

\begin{figure}[htpb]
\centering
\includegraphics[width=9cm,angle=0]{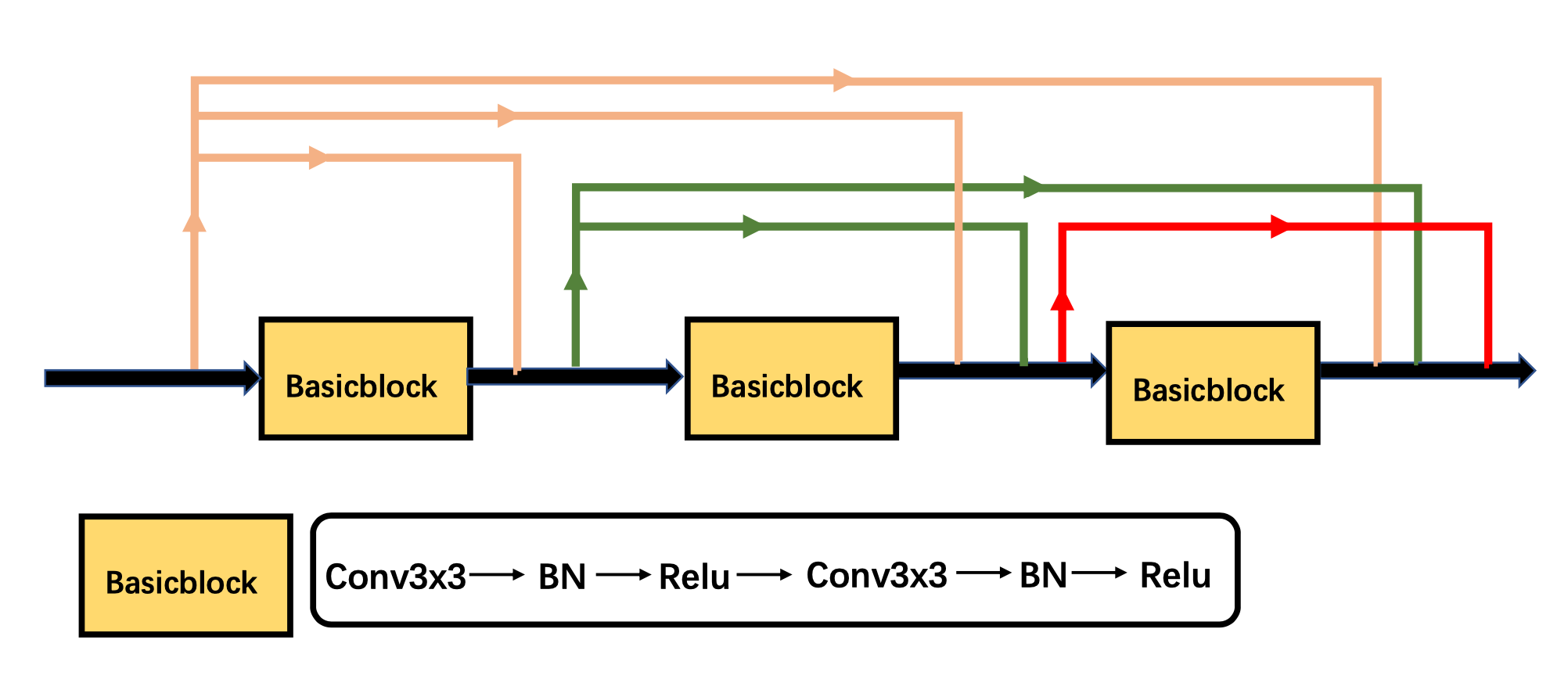}
\caption{The architecture of the Dense block. Each layer connects 
to every other layer in a feed-forward fashion.}
\label{f3}
\end{figure}

Inspired by the DAttNet \cite{Wu2020}, Unet \cite{Ronneberger2015}, Attention Unet \cite{Oktay2018} and DenseNet \cite{Huang2017}, we propose a framework V-DUnet. As illustrated in Fig. \ref{f2}, it is composed of two parts, the first part is the V part and the second part is the DUnet part. There are two inputs in V-DUnet, one is  the measurement image ${\bf y}$ with $256 \times 256$ pixels recorded by the CCD, the other is the reconstructed DGI result with size $128 \times 128 \times 15$. The input ${\bf y}$  is firstly reshaped into four channels with size $128 \times 128 \times 4$, then the reshaped result and DGI result pass through two convolutional block respectively and finally concatenated as one block (this process is corresponding to the V part of V-DUnet) and feeds into the DUnet part of V-DUnet. DUnet part is mainly designed based on DenseNet and Unet. DenseNet have four compelling advantages: (1) alleviate the vanishing-gradient problem, (2) strengthen feature propagation, (3) encourage feature reuse, and (4) substantially reduce the number of parameters \cite{Huang2017}. The Dense block used in V-DUnet is displayed in Fig.\ref{f3}.

\indent Additionally, we apply dropout layers to prevent overfitting \cite{Krizhevsky2017}, and batch normalization (BN) layers to speed up the convergence of loss function \cite{Ioffe2015}. The attention gate (AG) is also used to eliminate the irrelevant and noisy responses in Unet skip connections process, and enhance the salient features which pass through the skip connections \cite{Ronneberger2015, Oktay2018}. Here we introduce the FFDNet \cite{Zhang2018} in the training process as the denosing part of V-DUnet. It can deal with a wide range of noise levels and easily remove spatially variant noise by specifying a non-uniform noise level map with a single network.  

       \indent 
       The random sensing matrix $\Phi $ \cite{Wang2020, Zhang2019} and the structural similarity (SSIM) \cite{Wang2004, Zhang2020} between the ground truth and the reconstructed results are introduced into the loss function. Therefore, the loss function of our V-DUnet can be finally expressed as       
\begin{eqnarray}
  & Loss=&\alpha \| X-\hat {X}\|_1 + \beta \| Y - \Phi  \hat {X}\|_1 \nonumber\\
             &&  + \ \gamma [1-ssim(X, \hat {X})], 
 \label{eq2}
\end{eqnarray}  
here we set $\alpha=50$, $\beta=1$ and $\gamma=50$. $X$ represents the ground truth of the original HSI while $\hat {X}$ is the corresponding reconstructed HSI from the net. ssim($X,\hat {X}$) represents the SSIM between $X$ and $\hat {X}$, and it is formulated as
 
 \begin{table*}[htpb]
\caption{The average evaluation results on the ICVL, CAVE and Minho datasets. $225$ ICVL HSIs, $279$ CAVE HSIs and $201$ Minho HSIs are used to average evaluate PSNR, SSIM and SAM, respectively.}
\label{table1}
\renewcommand{\arraystretch}{1.8}
\centering
\begin{tabular}{|c|c|r|r|r|r|r|r|r|r|r|r|}
	\hline
	\multirow{2}{*}{Net}  &\multirow{2}{*}{Input}  &\multicolumn{3}{c|}{ICVL(225)} &\multicolumn{3}{c|}{CAVE(279)} &\multicolumn{3}{c|}{Minho(201)}	 \\\cline{3-11}  	           
	                                                  &&  PSNR  &   SSIM & SAM   &  PSNR  &   SSIM & SAM &  PSNR  &   SSIM & SAM       \\  \hline     
	    	\multirow{3}{*}{Unet}     &   only y                   & 19.5750    &  0.4791 & 0.3698  & 16.9264   &  0.4189 & 0.4939  & 17.9258    &  0.3917 & 0.4207 \\  \cline{2-11}       
		                                  &   only DGI                & 25.1347    &  0.7557 & 0.1793  & 21.5853   &  0.6683 & 0.3068  & 21.5046    &  0.6676 & 0.2739\\  \cline{2-11}
					            &   y+DGI            & 25.5148   &  0.7720 & 0.1707  & 21.7931    &  0.6789 & 0.3034  & 21.6336    &  0.6852 & 0.2676 \\\cline{1-11} 
	  	
	\multirow{3}{*}{Proposed}    &  only y                  & 20.9977    &  0.6002 & 0.2969  & 18.2602   &  0.5476 & 0.4119  & 19.0723    &  0.4986 & 0.3671  \\  \cline{2-11}
		                                 &   only DGI                & 25.7483    &  0.7635 & 0.1774 & 22.8264    &  0.7007 & 0.2919  & 22.8366    &  0.7037 & 0.2429 \\  \cline{2-11}
		                                 &   y+DGI           & \bf 26.9447    & \bf 0.7978 & \bf0.1565 & \bf23.4499    &  \bf0.7303 & \bf0.2799  & \bf23.1362    &  \bf0.7234 & \bf0.2403 \\   
        \hline                             
	                                   
\end{tabular}
\end{table*}

\begin{table*}[htpb]
\caption{Six different scenes reconstructed by different algorithms.}
\label{table3}
\renewcommand{\arraystretch}{1.8}
\centering
\begin{tabular}{|c|r|r|r|r|r|r|r|r|r|r|r|r|r|} 
	\hline
	\multirow{2}{*}{Algorithm}   &\multicolumn{3}{c|}{Ours} &\multicolumn{3}{c|}{PICHCS} &\multicolumn{3}{c|}{TwIST} &\multicolumn{3}{c|}{DGI}  	 \\ \cline{2-13}  	           
	                 &  PSNR  &   SSIM & SAM   &  PSNR  &   SSIM & SAM &  PSNR  &   SSIM & SAM    &  PSNR  &   SSIM & SAM   \\  \hline     
	    	\multirow{1}{*}{Scene 1}     
					             & \bf30.5125    &  \bf0.8827 & \bf0.1239 & 25.4607    &  0.5704 & 0.2668  & 20.3763    &  0.2691 & 0.3766 & 14.4801    &  0.3471 & 0.5073 \\   \cline{1-13} 
					             
                \multirow{1}{*}{Scene 2}   
                                                      &  \bf30.7070   & \bf0.9010 & \bf0.0969 & 24.8118    &  0.4440 & 0.2174 & 19.5310    &  0.1943 & 0.4372  & 14.8900    &  0.4787 & 0.2908\\   \cline{1-13}
                                                      
                   \multirow{1}{*}{Scene 3}  
                                                        &  \bf32.2708    & \bf0.8778 & \bf0.1659 & 25.4932   &  0.6471 & 0.3537 & 24.5597    &  0.4074 & 0.4046  & 14.8006   &  0.3795 & 0.4141 \\   \cline{1-13}                                      
                 
                  \multirow{1}{*}{Scene 4}  
                                                        &  \bf31.3115    & \bf0.8861 & \bf0.1897 & 25.9568    &  0.5729 & 0.3253  & 27.4102    &  0.6281 & 0.3992  & 12.2076    &  0.2326 & 0.5289\\   \cline{1-13}   
                   
                   \multirow{1}{*}{Scene 5}    
                                                       &  \bf32.2683    & \bf0.8678 & \bf0.1899 & 25.3419    &  0.4434 & 0.3360 & 23.7993   &  0.3571 & 0.4654 & 16.7134    &  0.5047 & 0.3587 \\   \cline{1-13} 
                    
                    \multirow{1}{*}{Scene 6}    
                                                        &  \bf31.1425    & \bf0.8922 & \bf0.1437 & 21.3948    &  0.4542 & 0.3170  & 20.5054    &  0.2671 & 0.5294 & 14.4138    &  0.4230 & 0.4168\\   \cline{1-13}    
                                                        
                   \multirow{1}{*}{Average}    		                                       
                                                        &  \bf31.3688   & \bf0.8846 & \bf0.1523 & 24.7432    &  0.5220 & 0.2961  & 22.6970   &  0.3538 & 0.4354 & 14.6509    &  0.3943 & 0.4194  \\   \cline{1-13}                         
                 
					             
             \hline    		                                   
\end{tabular}
\end{table*}

\begin{table*}
\caption{Anti-noise performance comparisons on the ICVL, CAVE and Minho datasets for the cases with SNR 30 dB and SNR 10 dB. 225 ICVL HSIs, 279 CAVE HSIs and 201 Minho HSIs are used to average evaluate PSNR, SSIM and SAM, respectively.}
\label{table2}
\renewcommand{\arraystretch}{1.8}
\centering
\begin{tabular}{|c|r|r|r|r|r|r|r|r|r|r|} 
	\hline
	\multirow{2}{*}{SNR}   &\multicolumn{3}{c|}{ICVL(225)} &\multicolumn{3}{c|}{CAVE(279)} &\multicolumn{3}{c|}{Minho(201)}	 \\ \cline{2-10}  	           
	                 &  PSNR  &   SSIM & SAM   &  PSNR  &   SSIM & SAM &  PSNR  &   SSIM & SAM       \\  \hline     
	    	\multirow{1}{*}{30 dB}     
					             & 26.9447    &  0.7978 & 0.1565 & 23.4499    &  0.7303 & 0.2799  & 23.1362    &  0.7234 & 0.2403 \\   \cline{1-10} 
					             
                \multirow{1}{*}{10 dB}    		                                       &  26.8888    & 0.7890 & 0.1526 & 23.2716    &  0.7157 & 0.2814  & 22.5408    &  0.7058 & 0.2421 \\   
                 
					             
             \hline    		                                   
\end{tabular}
\end{table*}

\begin{eqnarray}
ssim(X,\hat{X})=\frac{(2\bar{w}_{X}\bar{w}_{\hat{X}}+C_1)(2 \sigma_{{{w}_{X}}{w}_{\hat{X}}}+C_2)}{({\bar{w}_{X}}^2+{\bar{w}_{\hat{X}}}^2+C_1)({\sigma^2_{w_{X}}}+{\sigma^2_{w_{\hat{X}}}}+C_2)},
 \label{eq3}
\end{eqnarray} 
 where $w_X(w_{\hat{X}})$ represents the region of image $X$($\hat{X}$) within window $w$ while $\bar w_{X}(\bar w_{\hat{X}})$ is the mean of ${w_{X}}(w_{\hat{X}})$.  ${\sigma^2_{w_{X}}}({\sigma^2_{w_{\hat{X}}}})$ is the variance of $w_X(w_{\hat{X}})$,  $\sigma_{{{w}_{X}}{w}_{\hat{X}}}$ represents the co-variance between $w_X$ and $w_{\hat{X}}$. $C_1$ and $C_2$ are constants (experimentally set as $1 \times 10^{-4}$ and $9  \times10^{-4}$), the window $w$ is set to $11$ \cite{Zhang2020}.

\section{ Simulation Results}
Three public HSI datasets are used to evaluate our method , including the ICVL dataset \cite{Arad2016}, CAVE dataset \cite{Yasuma2010} and the Minho dataset \cite{Nascimento2002}. The ICVL dataset consists of $201$ HSIs ($1024 \times 1392 \times 31$) and the CAVE dataset consists of $32$ images ($512  \times 512 \times 31$), the spectral bands of both the ICVL and CAVE datasets are ranged from $400$ $nm$ to $700$ $nm$ with $10$ $nm$ intervals. The Minho dataset consists of $30$ HSIs ($820 \times 820 \times 31$), the wavelength range of $410$ $nm$$-$$720$ $nm$ was sampled at $10$ $nm$ intervals. We choose $15$ channels with spectral range from $560$ $nm$ to $700$ $nm$ in those datasets. 

\begin{figure*}[htpb]
\centering
\includegraphics[width=18cm,angle=0]{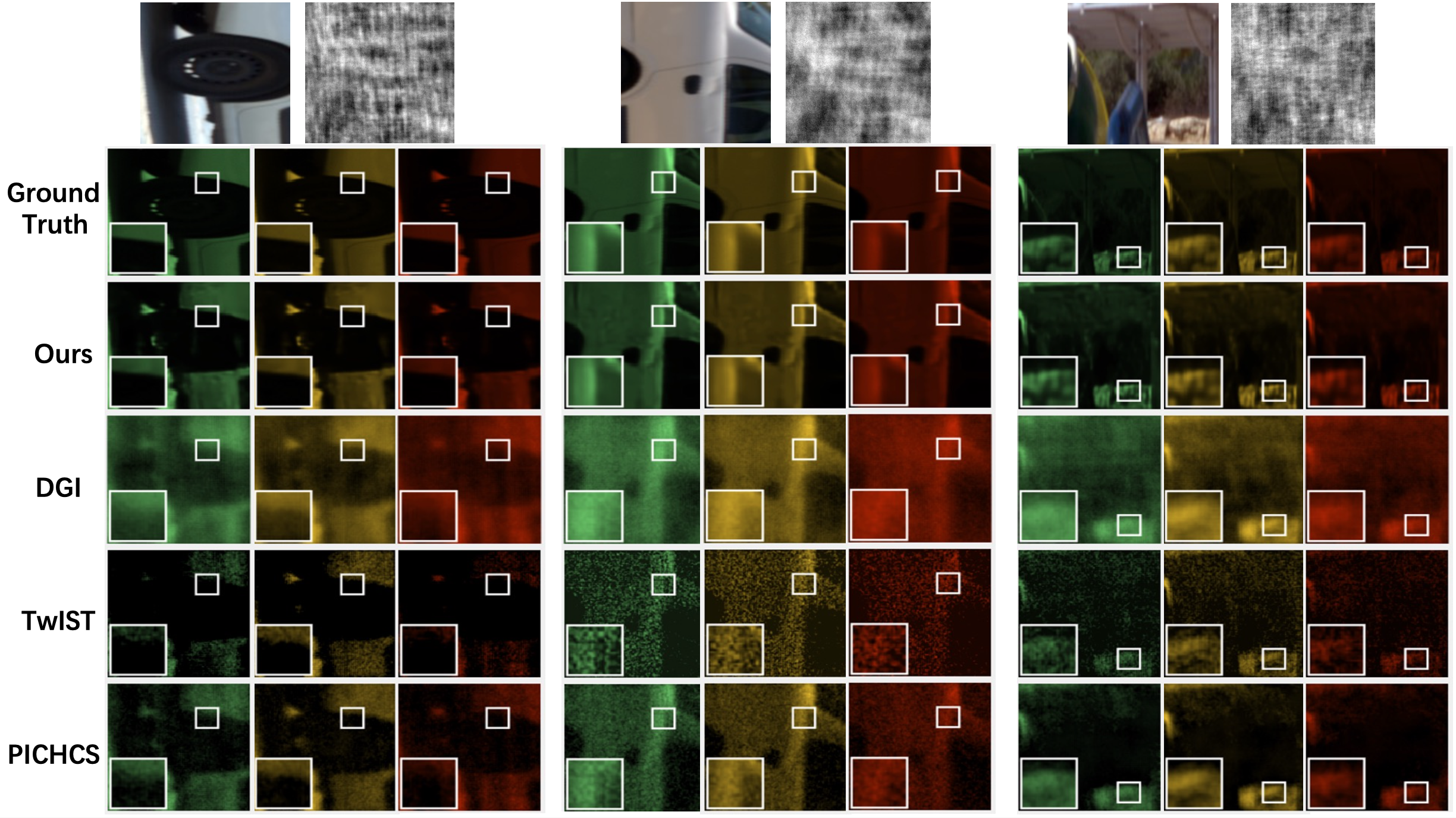}
\caption{Exemplar reconstructed images by $4$ algorithms for three scenes (from left to right: Scene $1$, Scene $2$, Scene $3$). The upper figures are the synthetic RGB and the image ${\bf y}$ respectively. Three ($560$ $nm$, $630$ $nm$ and $700$ $nm$) out of $15$ spectral channels are shown to compare with the ground truth.}
\label{f4}
\end{figure*} 

\begin{figure*}[htpb]
\centering
\includegraphics[width=18cm,angle=0]{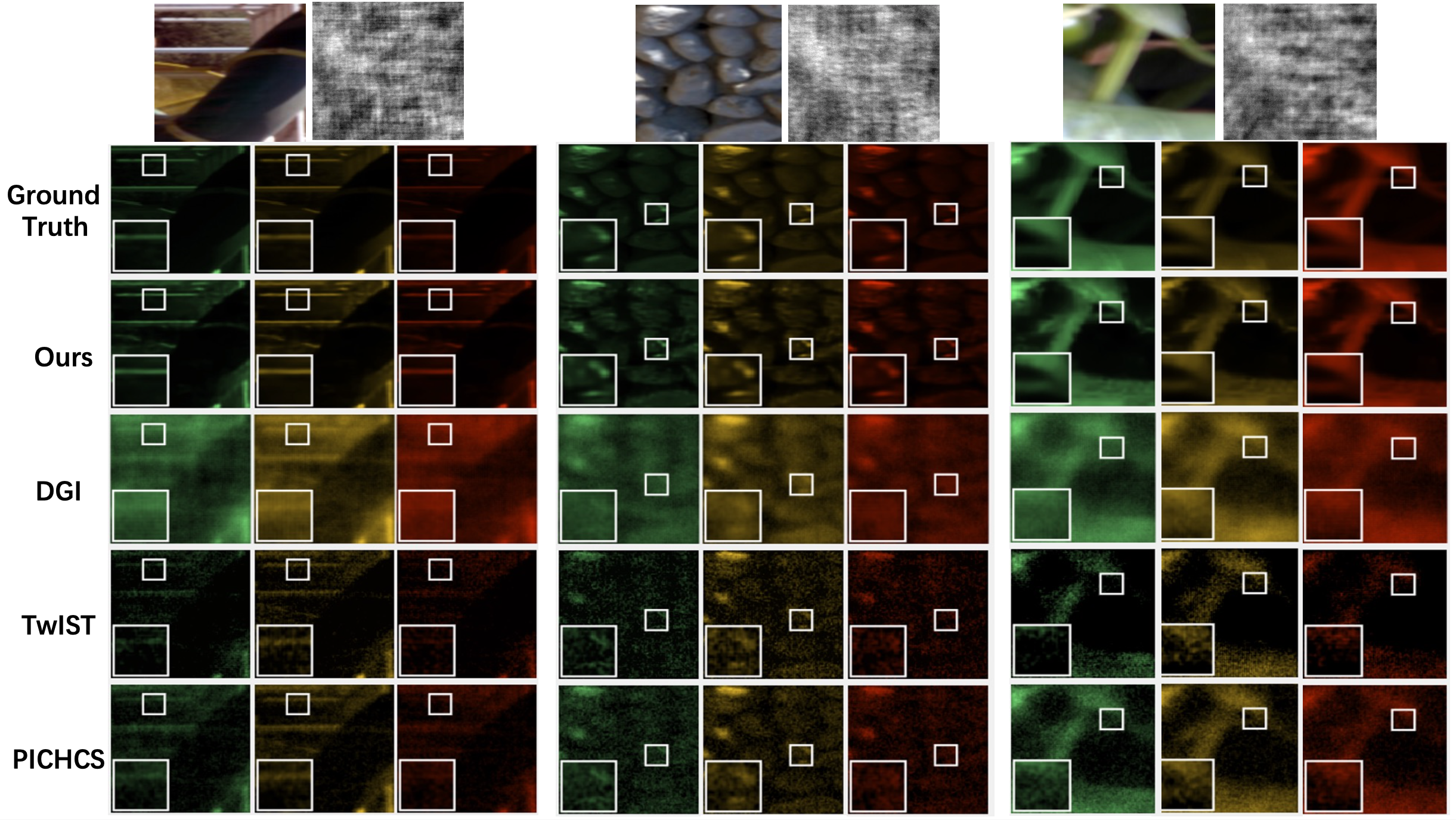}
\caption{Exemplar reconstructed images by $4$ algorithms for three scenes (from left to right: Scene $4$,Scene $5$, Scene $6$). The upper figures are the synthetic RGB and the image ${\bf y}$ respectively. Three ($560$ $nm$, $630$ $nm$ and $700$ $nm$) out of $15$ spectral channels are shown to compare with the ground truth.}
\label{f5}
\end{figure*}

\indent To eliminate the overfitting effect, we manually exclude $91$ HSIs with similar background or contents and selected $110$ HSIs in ICVL dataset. Then we randomly select $101$ HSIs in the subsets for training and thus use the rest $9$ HSIs for testing. To formulate the training and validation datasets, HSI patches with the size of $128 \times 128 \times 15$ are uniformly extracted with the stride of $128$ from the above $101$ HSIs in ICVL dataset. We randomly select $90\%$ patches for training and $10\%$ patches for validation. As for the CAVE and Minho dataset, none of them has been included in the training dataset, they are only used for testing. We randomly crop $225$ HSI patches from the rest $9$ HSIs in ICVL dataset, $279$ HSI patches from the CAVE dataset and $201$ HSI patches from the Minho dataset for testing. All the models are only trained on ICVL dataset and anther input ${\bf y}$ for training is obtained by Eq.\ref{eq1}, where the detected Signal to Noise Ratio (SNR) is 30 dB and $\Phi$ is obtained by the calibration of GISC spectral camera. 
\begin{figure}[htpb]
\centering
\includegraphics[width=9cm,angle=0]{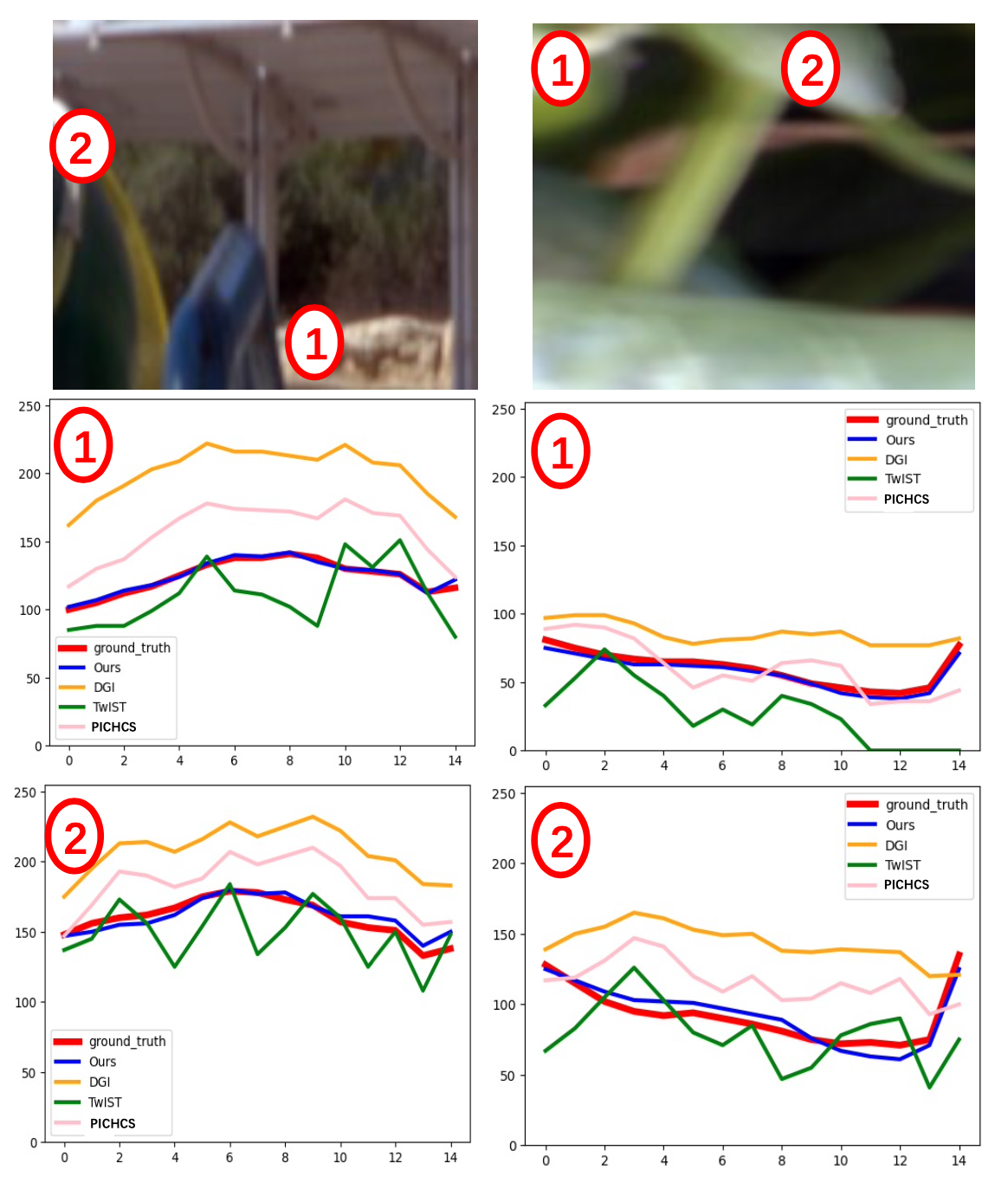}
\caption{Spectral curves of the Scene 3 and Scene 6.}
\label{f6}
\end{figure}

\indent Three quantitative image quality metrics, including peak signal-to-noise ratio (PSNR), SSIM and spectral angle mapping (SAM) \cite{Kruse1993}, are used to evaluate the performance of all methods. Larger PSNR, SSIM and the smaller SAM values suggest better reconstruction performance, and vice versa. 
       
        \indent     The effects of different inputs of the V part in the net have also taken into account during the net design process, see TABLE \ref{table1}. It is obvious that when only ${\bf y}$ is used as input, the net reconstruction result is unsatisfactory for neither Unet nor DUnet. When the net inputs are DGI and ${\bf y}$, the average improvement in PSNR of reconstructed result has greatly achieved about 6 dB compared with the case when the net input is only ${\bf y}$, and about 1 dB compared with the case when the net input is only DGI. As shown in TABLE \ref{table1}, compared with the case when only basic Unet is used in the second part of the net, DUnet which is mainly designed by Dense block and Unet obtains better reconstruction performance. 
   
       To verify the performance of our proposed method, we compare it with several representative reconstruction methods including DGI, TwIST \cite{BioucasDias2007}, and PICHCS \cite{Shiyu2015}. We have made great effort to achieve the best results of all those competitive methods. To visualize the experimental results for all methods, several representative reconstructed image for 6 scenes on ICVL dataset are shown in Fig. \ref{f4} and Fig. \ref{f5}. The PSNR, SSIM and SAM using V-DUnet and other three algorithms are listed in TABLE \ref{table3}. Fig. \ref{f4} and Fig. \ref{f5} shows that our V-DUnet has achieved visually pleasant results with more details of the images compared with other three methods, which is consistent with the numerical evaluation metrics listed in TABLE \ref{table3}. The spectral curves of the reconstruction and ground truth have been plotted in Fig. \ref{f6}. It can be seen that spectral curves of our method are more close to the ground truth which further demonstrates that V-DUnet can extract more spectral information compared with other methods. TABLE \ref{table2} shows the noise tolerance performance of V-DUnet, where the cases with SNR 30 dB and 10 dB are verified with the same training weights. We can see that when the SNR decreases to $10$ dB from $30$ dB, the reconstructed results just slightly degenerated, which demonstrates that our method is robust to the noise.  
       

\section{Conclusion} 

This paper aims to improve the image reconstruction quality and real-time performance in GISC spectral camera. Inspired by the recent advances of deep learning, we proposed an end-to-end V-DUnet to obtain the 3D hyperspectral images in GISC spectral camera. It can quickly reconstruct high-quality 3D hyperspectral images by integrating DenseNet into the Unet framework and setting both differential ghost imaging results and the detected measurements as the network’s input. As observed in TABLE \ref{table2}, V-DUnet is also robust to the noise. In view of the well performance of the network, it is expected to be applied into super-resolution imaging via discernibility in high-dimensional light-field space \cite{Tong2020} and other high-dimensional imaging system \cite{Chu2021, Arce2014}.





%





\ifCLASSOPTIONcaptionsoff
  \newpage
\fi



\bibliographystyle{IEEEtran}
\bibliography{IEEEabrv,V-DUnet}

%








\end{document}